# The Shackles of Peer Review: Unveiling the Flaws in the Ivory Tower

LIU Ying[1], YANG Kaiqi[2], LIU Yue[1], DREW Michael G. B[3]

[1]College of Chemistry and Chemical Engineering, Shenyang Normal University, Shenyang, P. R. China,110034, yingliusd@163.com (Ying Liu), yueliusd@163.com (Yue Liu)

[2]Dalian Bosi Middle School, Dalian Liaoning, 1263078910@qq.com

[3]School of Chemistry, The University of Reading, Whiteknights, Reading RG6 6AD, UK, m.g.b.drew@reading.ac.uk

Abstract:

This essay delves into the ethical dilemmas encountered within the academic peer review process and investigates the prevailing deficiencies in this system. It highlights how established scholars often adhere to mainstream theories not out of genuine belief, but to safeguard their own reputations. This practice perpetuates intellectual conformity, fuels confirmation bias, and stifles dissenting voices. Furthermore, as the number of incorrect papers published by influential scientists increases, it inadvertently encourages more researchers to follow suit, tacitly endorsing incorrect viewpoints. By examining historical instances of suppressed ideas later proven valuable, this essay calls for a reevaluation of academia's commitment to genuine innovation and progress which is usually achieved by applications of fundamental principles in from textbooks.

# 同行评审伦理：象牙塔中的缺陷

刘颖 [1]，杨凯奇 [2]，刘跃 [1]，DREW Michael G. B.[3]

[1]沈阳师范大学化学与化工学院，中国辽宁省沈阳市，110034，电子邮件：yingliusd@163.com（通讯作者：刘颖），yueliusd@163.com（刘跃）

2 大连博思中学，辽宁大连，1263078910@qq.com

[3]里丁大学化学学院，英国 Reading RG6 6AD，电子邮件：m.g.b.drew@reading.ac.uk

摘要：

本文深入探讨了学术同行评审过程中遇到的伦理难题，并研究了该体系中普遍存在的缺陷。文章强调，学术权威通常坚守主流理论，不是因为他们真正相信这些理论，而是为了维护自己的声

誉。这种做法导致学术欺凌，助长从众心理，抑制非主流声音，阻碍科学进步。此外，随着有影响力的科学家大量发表错误的论文以及研究人员效仿，更多的人成为主流理论的既得利益者，使得错误理论更难纠正。通过审视历史上被打压，后来被证明推动了科学进步的思想实例，本文呼吁学术界建立对真正创新和进步的宽松环境，因为真正的深入研究往往是应用教科书中的基本原理的结果。

关键词：同行评审，主流理论，主流学术权威，学术争鸣，理论纠错，科学进步。

## Introduction

Peer review has long been regarded as the cornerstone of academic rigor, a process designed to sift the wheat from the chaff in the world of scholarly research. However, beneath this veneer of impartiality lies a deeply ingrained flaw: the resistance of established scholars to critique the mainstream theories they have long championed. Peer review can be slow, subjective, and prone to bias. It is not effectively to detect errors or fraud. [1, 2, 3, 4, 5] This essay focuses on the academic ethics of peer review and explores the prevalent problem of scholars upholding incorrect theories not due to genuine conviction but rather because their reputations are closely linked to these theories. It also highlights how this practice spawns a cycle where more researchers endorse these incorrect views, suppressing genuine innovation and progress. The academic community's blind adherence to the status quo reveals a disturbing underbelly of academic dogma and stifling intellectual progress. By referencing historical examples of dismissed ideas that later proved invaluable, this essay calls for a reevaluation of academic norms.

## The Herd Mentality

The academic world, like many other fields, can be resistant to change and often favors established conclusions and methodologies. Scientific paradigms and consensus can persist for extended periods, even when they contain flaws or inaccuracies. This is due to the resistance of established scholars to challenge prevailing viewpoints. This resistance can lead to skepticism and slow acceptance of innovative or unconventional ideas.

Truly groundbreaking research often challenges existing conclusions and may face resistance from reviewers and established scholars. However, it is through such research that science progresses and evolves. The academic community can sometimes resemble a hierarchical structure where established authorities hold significant influence. Challenging these authorities can be difficult for unconventional scholars. The peer review system perpetuates itself through a cycle of self-preservation. Scholars who have invested decades in promoting a particular theory are unlikely to admit its flaws, even when confronted with compelling evidence to the contrary. To do so would be to admit that their life's work may have been misguided. This not only threatens their ego but also their livelihood, as academic positions and funding often depend on maintaining a favorable reputation. However, it poses a significant threat to academic progress as scholars prioritize self-preservation over truth.

In academia, reputation functions as the pivotal factor that can make or break a scholar's career. Established scholars who have risen to prominence through their work on mainstream theories often find themselves in a catch-22 situation. On one hand, their prestige is built upon the very theories they have dedicated years to studying, and on the other, they possess a keen awareness of the flaws inherent in these theories, especially when they have been exposed to new theories that diverge from the mainstream. This predicament places them in a difficult position, as acknowledging these flaws can potentially jeopardize their careers and perpetuate a culture of intellectual integrity deviate from conformity, which is detrimental to their authority. This conundrum creates a pernicious incentive structure that dissuades scholars from speaking out against the established order, as admitting their mistakes could jeopardize their careers and disrupt the culture of conformity they have grown accustomed to.

In a certain sense, the academic community resembles a faction-ridden "martial arts world," where academic authorities wield power akin to "sect leaders," and ordinary scholars lack the strength to challenge their viewpoints. As influential scientists publish erroneous papers, more researchers follow suit, perpetuating incorrect viewpoints. As the number of erroneous papers being published increases and more researchers follow the trend, everyone becomes a beneficiary, tacitly allowing these incorrect viewpoints to continue propagating. [6] There is a growing concern that the pressure for career advancement and the tendency to follow trends are leading to the publication of erroneous or low-quality papers, potentially perpetuating widely accepted but incorrect viewpoints.

The peer review process, while intended to be impartial, is susceptible to the herd mentality. Scholars reviewing manuscripts are often well-acquainted with the work of their peers, and they may hesitate to challenge widely accepted theories, fearing backlash from their colleagues. This herd mentality leads to a collective blindness to alternative perspectives, stifling innovation and intellectual diversity.

In the realm of science, conformity often prevails over dissent. A considerable number of individuals are inclined to align with authority, seemingly believing that this aligning somehow validates their own greatness. [7] Dissenting voices are often marginalized within academia, labeled as contrarians or heretics. This suppression of dissenting views hinders the critical examination of mainstream theories and discourages scholars from challenging the orthodoxy. As a result, promising alternative theories are left in the shadows, and intellectual progress is stifled.

History serves as the ultimate arbiter. It consistently reveals that what is often overemphasized by the prominent scholars of an era is often merely the intentional promotion of mediocrity, while that ideas suppressed or dismissed in their time can later prove to be of great value.[8] Scientific breakthroughs often hinge on unconventional scientists [9, 10, 11], referred to as "民科" or "non-mainstream scientists" in Chinese, and these terms can be translated into English as "pseudo-scientists" or "unconventional scientists." In the field of science, it is frequently these minority of "民科" that leads the way to groundbreaking discoveries. Innovation typically involves ideas that encounter resistance from the majority. True innovation transcends trends, as pursuing trends doesn't constitute genuine innovation. What garners agreement from the majority doesn't typically qualify as innovation.

The term "innovation" inherently implies deviation from the norm, while commonly accepted ideas usually represent established knowledge. The quest for broad consensus seldom fosters innovation, and following popular topics rarely nurtures creativity. Genuine innovation often thrives within specialized niches, where even the most widely cited papers may sometimes fall short of genuine novelty. History is replete with examples of ideas that were suppressed or dismissed in their time, only to later prove to be invaluable. For instance, the heliocentric model proposed by Copernicus challenged the geocentric worldview and was initially rejected but ultimately revolutionized astronomy. [12, 13]

Truly innovative concepts often come to light in the blogosphere rather than being confined to mainstream media. Hidden gems may lie within unnoticed preprints, waiting to be discovered. It highlights the limitations of contemporary thinking and the importance of revisiting ideas that were marginalized. Preprints ought to function as the principal medium for disseminating ideas that deviate from mainstream theories, igniting fresh perspectives and potentially catalyzing substantial advancements in science. However, preprint platforms and journals often tend to adopt the same criteria. If the evidence of the deviated ideas are dismissed from publication solely due to insufficient evidence, how can we ever accumulate such evidence?"

Confirmation bias is a well-documented psychological phenomenon in which individuals tend to seek out and interpret information in a way that confirms their preexisting beliefs. In the context of peer review, scholars may unconsciously cherry-pick evidence that supports the mainstream theory they champion, while dismissing or downplaying contradictory evidence. This bias further entrenches the flawed status quo.

## Obstacles Encountered by Innovative Breakthroughs in Peer Review

What prompted the initial lack of acceptance for inventions? What is the reason for the absence of major breakthroughs or discoveries in science in recent decades?

"Professor Braben argues that the introduction in the 1970s of the (peer) review of research proposals has led to a dearth of big scientific discoveries. The most radical ideas, he says, are unlikely to get funded because it is difficult to impress peers before they have been proven. … It (peer review) works well enough in the mainstream but it is at the margins where major discoveries are made, where people don't believe in the current wisdom and want to head off into dramatically different directions. To submit those ideas to peer review is disastrous". [3, 14]

This can also be traced back to the predominant emphasis in contemporary science on surface-level experimental research heavily reliant on cutting-edge instrumentation. This is a departure from the era of Newton, where scientists placed a strong emphasis on profound theoretical investigations employing mathematics to unveil the true essence underlying experimental phenomena. Deep research often relies on established textbook principles.

Often, authors are true experts, while referees only offer an external perspective and suggestions to enhance the manuscript. "So we have little evidence on the effectiveness of peer review, but we have considerable evidence on its defects. In addition to being poor at detecting gross defects and almost useless for detecting fraud it is slow, expensive, profligate of academic time, highly subjective, something of a lottery, prone to bias, and easily abused." [2]

When delving into history, it becomes apparent that groundbreaking original innovations faced skepticism and opposition within the industry upon their initial emergence, nearly teetering on the brink of extinction. [15] This phenomenon is not an isolated occurrence. The greater the innovation, the more it challenges and disrupts established norms, resulting in heightened resistance from traditional forces. [16]

A notable example can be found in the realm of microwave absorption, where the mainstream theoretical framework has been fundamentally flawed for a very long extended period. [17] Interestingly, this erroneous theory can be rectified using principles at college level, coupled with mathematical skills no more advanced than what is taught in junior middle school. However, there has been significant resistance to making this correction. [18] The correction has been disseminated through more than 20 publications, each presenting different perspectives, in reputable journals. These publications have garnered significant attention, evident from the substantial number of views and downloads. [19 - 21] Strikingly, there has been a conspicuous absence of negative comments or letters challenging these corrections. What's particularly surprising is that the erroneous mainstream theory continues to persist in practice, with numerous publications appearing in various journals, and almost all of them fail to address opposing viewpoints that contradict their conclusions. This omission can be seen as a form of intellectual dishonesty on the part of researchers who intentionally neglect to mention contrasting perspectives in relation to their findings. The misconduct becomes even more egregious when reviewers actively encourage others to publish incorrect papers in an attempt to dilute the impact of their own prior erroneous publications. It's worth mentioning that a very recent paper from another research group has confirmed the new theory regarding wave cancellation. [22]

Despite many papers exposing the flaws in the current microwave absorption theory from various angles, progress in this subject still faces significant hurdles in terms of publication. The reasons for the manuscript rejection are not based on critiques of the manuscript's arguments but rather on the notion that overturning established theories lacks the significance required for publication. Some reviewers suggest that the research only involving junior school mathematics is better suited for mathematical journals rather than material and physical ones. [18,23] Moreover, introducing a new absorption mechanism to supplant the erroneous previous one is often perceived as lacking the contribution of genuinely novel scientific insights within the manuscript.

Evidently, these reviewers may not have thoroughly examined the manuscripts. This becomes especially evident when the mathematical complexity of the manuscript is no more advanced than what is typically covered in junior middle school education, and the sole theoretical foundation employed is grounded in basic principles of general physics aimed at challenging the accepted theory. It is possible that many of the reviewers struggle to comprehend the arguments, even though these arguments are grounded in fundamentals not exceeding a college-level understanding. It appears that while many researchers have encountered these basic principles during their college education, they may not possess a deep understanding of them. This situation reflects a prevailing tendency in contemporary research, where a focus on superficial experiments overshadows the importance of fundamental principles found in textbooks. In-depth research often emerges from the application of fundamental scientific principles rather than from conducting superficial experiments by selectively cherry-picking data to support mainstream theories.

The established theory has held a dominant position in this contemporary microwave absorption research field for an extended period, accompanied by plenty of experimental reports "supporting" it. Thus, we cannot dismiss the significance of the opposite theory solely due to the apparent simplicity of the principles employed to overturn this dominant mainstream theory. It is noticed that the most groundbreaking manuscripts are frequently rejected under the pretext of lacking novelty. Manuscripts are frequently rejected because of the time-consuming process of reviewing subjects that challenge established theories. Nevertheless, journals and reprint platforms serve as arenas where diverse viewpoints collide, rendering the time factor less of a concern.

More reasonable feedbacks from reviewers regarding the rejections of the relevant manuscripts center on that abundance of reports support the existing theory, while only a few originate from the single research group of the manuscripts that supports the opposing theory. It is understandable that there are more reports affirming the established theory in comparison to the newer one. However, if new evidence contradicting the current theory is not allowed to be published, how can such evidence ever accumulate? Furthermore, the question arises as to how much evidence is required to be enough to overturn an accepted theory. Mathematically, proving a theorem can be a challenging endeavor, but overturning one only necessitates a single piece of opposing evidence. The manuscripts astutely highlight that the data often cited as supportive of this accepted theory actually contradict it. In the case of the manuscript's arguments a wealth of apparent evidence against the current theory have been provided, which has often been dismissed by the researchers solely because the current theory cannot accommodate it. Intriguingly, many reviewers who opted to disregard these pertinent arguments were quick to reject the manuscripts without giving the authors an opportunity to defend their work, even though journals should serve as forums where diverse viewpoints can collide.

In cases where a theory undergoes theoretical invalidation, any experiments conducted within the framework of that theory are destined to falter, mirroring the futility of attempting to design a perpetual machine once it has been refuted by the fundamental tenets of thermodynamics. It worth noting the nations that "Can so many scientists have been wrong over the eighty years since 1925? Unhappily, yes. The mainstream in science, as any scientist will tell you, is often wrong. Otherwise, come to think of it, science would be complete. Few scientists would make that claim, or would want to … Scientists are often tardy in fixing basic flaws in their sciences despite the presence of better alternatives" [24], "Now pretty much every journal uses outside experts to vet papers, and papers that don't please reviewers get rejected … Weak-link thinking makes scientific censorship seem reasonable, but all censorship does is make old ideas harder to defeat. Remember that it used to be obviously true that the Earth is the center of the universe, and if scientific journals had existed in Copernicus' time, geocentrist reviewers would have rejected his paper and patted themselves on the back for preventing the spread of misinformation. ... We still don't understand basic truths about the universe, and many ideas we believe today will one day be debunked ", [12], "A new scientific truth does not triumph by convincing its opponents and making them see the light, but rather because its opponents eventually die, and a new generation grows up that is familiar with it", [25] and "some scientists wondered how a questionable line of research persisted for so long … experts were just too timid to take a stand." [26].

Curiously, when the manuscripts eventually get published, not a single referee from the earlier journals, who vehemently opposed their publication, steps forward to write a comment letter pointing out supposed "flaws" in these opposite papers. It appears that some reviewers can be rather cavalier with their comments in the role of reviewers but seem to recognize the weight of responsibility when they're writing a public comment letter.

If the comments of the reviewers are explicitly and pragmatically articulated, their primary purpose should be to highlight the deficiencies within the authors' arguments. This, in turn, would enable the referees to reasonably infer that the authors possess an incomplete understanding of the established theory. In instances where referees struggle to identify these limitations within the manuscript's arguments, it may imply a steadfast allegiance to the outdated theory and a reluctance to embrace a more open-minded perspective. Established theories tend to dominate the discourse, while the newly emerging ones often face misunderstanding. In the field of science, groundbreaking results are typically initially discovered by a minority of researchers.

What is particularly noteworthy is the academic community's propensity to disregard instances of academic misconduct during the peer review process, often displaying a notably higher level of tolerance and occasionally even indulgence, in contrast to cases involving image misuse and data fabrication. The issues stemming from the rejection of valuable ideas due to academic misconduct in peer review often persist without resolution until a new generation takes it upon themselves to address and rectify these concerns, typically after the influential generation of established scholars has passed away. [25]

## Reverting to the Authentic Science of Newton's Era

In the modern age, academic journals are classified into different categories, creating an ironic situation where publishing subpar articles in top-tier journals is often celebrated as groundbreaking. Additionally, the extensive use of promotions and advertisements has allowed mediocre articles to overshadow genuinely innovative ones. Science has undergone a transformation into a profession, and academia has taken on the appearance of a scholarly game. Academic authorities actively encourage mainstream scientists to engage in this scholarly pursuit. However, if we were to eliminate journal rankings and various promotional mechanisms, we might witness a return to the true science of Newton's Era, the way it should be. In such a scenario, mediocre articles would naturally fade into obscurity, while valuable contributions would withstand the test of time and receive the recognition they deserve from history.

The article primarily explores the ethical challenges that exist within the academic peer review process, particularly when it pertains to manuscripts presenting viewpoints contrary to the mainstream theory. These concerns are directly linked to the progress of scientific knowledge and hold more significant implications than instances of academic misconduct such as image misuse and data fabrication, which may occur within the mainstream academic framework.

It's important to note that while these challenges exist, the scientific and academic communities continue to evolve. Efforts should be made to improve peer review processes, encourage openness to new ideas, and promote transparency in research. Science, in the long run, remains a self-correcting process, and the pursuit of truth and knowledge continues to drive progress despite the obstacles.

## Conclusions

This essay centers its focus on ethical considerations within the realm of academic peer review. In its ideal form, peer review should serve as a bastion of intellectual integrity, nurturing an environment where theories are rigorously tested and challenged. Nevertheless, the flaws in this system become evident when established scholars, who are aware of the weaknesses in mainstream theories, choose to uphold them for the sake of self-preservation. This practice perpetuates a culture characterized by conformity, confirmation bias, and the suppression of dissent.

To truly propel human knowledge forward and promote intellectual progress, academia must confront and rectify these shortcomings within the peer review process. It is imperative that scholars have the freedom and encouragement to critically assess prevailing theories, even when doing so challenge their own reputations. Only through this approach can we hope to break free from the shackles of intellectual stagnation and pave the way for genuine innovation and discovery in the academic world.

To promote genuine innovation and progress, academia should address the flaws in peer review, where scholars cling to mainstream theories to safeguard their reputations. This practice encourages conformity, reinforces confirmation bias, and suppresses dissent. Furthermore, the proliferation of flawed papers tempts more researchers to follow suit, tacitly endorsing incorrect viewpoints. By acknowledging historical instances of undervalued ideas, academia can break free from intellectual stagnation and cultivate a culture that values truth over reputation.